\documentclass[superscriptaddress,twocolumn,amsmath, nofootinbib]{revtex4}
\usepackage{amsfonts}
\usepackage{graphicx}
\usepackage[latin1]{inputenc}
\usepackage{amsmath}
\usepackage{amssymb}

\begin{document}
%\draft

%\preprint{APS/123-QED}

\title{World-line geometry probed by fast spinning particle.}

\author{Alexei A. Deriglazov }
\email{alexei.deriglazov@ufjf.edu.br} \affiliation{Depto. de Matem\'atica, ICE, Universidade Federal de Juiz de Fora,
MG, Brasil} \affiliation{Laboratory of Mathematical Physics, Tomsk Polytechnic University, 634050 Tomsk, Lenin Ave. 30,
Russian Federation}
\author{Walberto Guzm\'an Ram\'irez }
\email{wguzman@cbpf.br} \affiliation{Depto. de Matem\'atica, ICE, Universidade Federal de Juiz de Fora, MG, Brasil}

\date{\today}% It is always \today, today,

\begin{abstract}
Interaction of spin with electromagnetic field yields an effective metric along the world line of spinning particle. If
we insist to preserve the usual special-relativity definitions of time and distance, critical speed which spinning
particle can not overcome during its evolution in external field differs from the speed of light. Instead, we can
follow the general-relativity prescription to define physical time and distance in the presence of electromagnetic field.
With these definitions, critical speed coincides with the speed of light. Effective metric arises also when spin interacts with gravitational field.
\end{abstract}

%\pacs{11.10.Ef, 03.65.Ca}% PACS, the Physics and Astronomy
%\keywords{Semiclassical Description of Relativistic Spin, Dirac Equation, Theories with Constraints}%Use showkeys class option if keyword
\maketitle

%\maketitle %\noindent
%{\bf DOI:}
%{\bf PACS numbers:} 11.10.Ef, 03.65.Ca \\
%{\bf Keywords:} Semiclassical Description of Relativistic Spin.

Typical relativistic equations of motion (EM) have singularity at some value of a particle
velocity. For instance, the standard Lagrangian of spinless particle in electromagnetic field
\begin{eqnarray}\label{FF.0}
L=\frac{1}{2\lambda}\dot x^2-\frac{\lambda}{2}m^2c^2+\frac{e}{c}A\dot x,
\end{eqnarray}
implies the equations
%\begin{eqnarray}\label{FF.1}
$\frac{d}{d\tau}\left[\frac{\dot x^\mu}{\sqrt{-\dot x^2}}\right]=\frac{e}{mc^2}F^\mu{}_\nu\dot x^\nu$,
%\end{eqnarray}
which became singular as $\dot x^2\rightarrow 0$. Using reparametrization invariance of EM, we take physical time as
the parameter, $\tau=t$, then $x^\mu=(ct, {\bf x}(t))$, $\dot x^\mu=(c, {\bf v}(t))$ and $\frac{1}{\sqrt{-\dot
x^2}}=\frac{1}{\sqrt{c^2-{\bf v}^2}}$.  For the acceleration ${\bf a}=\frac{d{\bf v}}{dt}$, the EM imply ${\bf v}{\bf
a}=[c^2-{\bf v}^2]^{\frac{3}{2}}\frac{e({\bf v}{\bf E})}{mc^3}$, that is the longitudinal acceleration vanishes as
$|{\bf v}|\rightarrow c$. Hence the singularity implies that the particles speed can not exceed the value $c$.

We will use the following terminology. The speed $v_{cr}$ that a particle can not exceed during its evolution in an
external field is called critical speed. The observer independent scale $c$ of special relativity is called, as usual, the speed
of light. According to the expression for ${\bf a}$ above, critical speed of a spinless particle coincides with the
speed of light.

Below we show that this situation changes in special relativity of spinning matter: if we preserve the usual
special-relativity definitions of time and distance, the speed of light does not represent special point of manifestly
relativistic EM of a particle with spin.

We consider the Lagrangian formulation of Frenkel electron \cite{Frenkel2} on an arbitrary electromagnetic background
which has been obtained in \cite{deriglazov2014Monster}. We demonstrate that interaction of Frenkel spin-tensor with
electromagnetic field necessarily modifies the factor $\frac{1}{\sqrt{-\dot x^\mu\eta_{\mu\nu}\dot x^\nu}}$, this
replaced by $\frac{1}{\sqrt{-\dot x^\mu g_{\mu\nu}\dot x^\nu}}$, where spin/field-dependent matrix $g_{\mu\nu}$
appeared instead of the Minkowski metric $\eta_{\mu\nu}$. The critical speed is determined now from the equation $\dot
xg\dot x=0$. If we insist on usual special-relativity definitions of time and distance, then even for rather simple
electromagnetic configurations this equation yields the critical speed a little more than the speed of light.

Next we suggest a possibility to maintain the equality $v_{cr}=c$. This can be achieved supposing that infinitesimal intervals of
physical time and distance in electromagnetic field should be defined according to general-relativity
prescription, that is in a space-time with the effective metric $g_{\mu\nu}$.

%Similar ideas were expressed earlier in a number of papers.
%Similar ideas have been considered previously in several
%papers.

Similar ideas were mentioned before in a number of works. The appearance of trajectories with space-like four-velocity
was remarked by Hanson and Regge in their model of spherical top in electromagnetic field
\cite{hanson1974relativistic}. Space-like trajectories of the model in gravitational fields were studied by Hojman,
Regge \cite{Hojman:1976} and by Hojman, Asenjo \cite{Hojman:2013}. A possibility of deformed relation between proper
and laboratory time in the presence of electromagnetic field was discussed by van Holten in his model of spin
\cite{Holten}.

Consider spinning particle with mass $m$, electric charge $e$ and magnetic moment $\mu$
interacting with an arbitrary electromagnetic field $F_{\mu\nu}=\partial_\mu A_\nu-\partial_\nu A_\mu$. The manifestly
Poincare and reparametrization invariant Lagrangian on configuration space with coordinates $x^\mu(\tau)$,
$\omega^\mu(\tau)$ and $\lambda(\tau)$ reads \cite{deriglazov2014Monster}
\begin{eqnarray}\label{FF.1}
L=\frac{1}{4\lambda}\left[\dot xN\dot x+\lambda D\omega ND\omega-\right. \qquad \qquad \cr \left.\sqrt{\left[\dot xN\dot x+\lambda D\omega
ND\omega\right]^2-4\lambda(\dot xND\omega)^2}\right]- \cr \frac{\lambda}{2}m^2c^2+\frac{\alpha}{2\omega^2}+\frac{e}{c}A\dot x.
\qquad \qquad
\end{eqnarray}
This depends on one free parameter $\alpha=\frac{3\hbar^2}{4}$, this particular value corresponds to spin one-half
particle. Similarly to Eq. (\ref{FF.0}), the only auxiliary variable is $\lambda$, this provides the mass-shell
condition. It has been denoted
%\begin{equation}\label{m1.1}
$N^{\mu\nu}\equiv\eta^{\mu\nu}-\frac{\omega^\mu\omega^\nu}{\omega^2}$, then $N^{\mu\nu}\omega_\nu=0$.
%\end{equation}
To introduce coupling of the position variable $x$ with electromagnetic field, we have added the minimal interaction
$\frac{e}{c}A_\mu\dot x^\mu$. As for basic variables of spin $\omega^\mu$, they couple with $A^\mu$ through the term
%\begin{eqnarray}\label{m.2}
$D\omega^\mu\equiv\dot\omega^\mu-\lambda\frac{e\mu}{c}F^{\mu\nu}\omega_\nu$.
%\end{eqnarray}
The Frenkel spin-tensor is a composite quantity constructed from $\omega^\mu$ and its conjugated momentum $\pi^\mu$ as
follows\footnote{The basic variables $\omega^\mu$ do not represent observable quantities since they are not invariant
under spin-plane local symmetry presented in the model \cite{DPM2}.}:
\begin{eqnarray}\label{FF.2}
J^{\mu\nu}=2(\omega^\mu\pi^\nu-\omega^\nu\pi^\mu)=(J^{i0}=D^i, ~ J_{ij}=2\epsilon_{ijk}S_k).
\end{eqnarray}
Here $S_i$ is three-dimensional spin-vector and $D_i$ is dipole electric moment \cite{ba1}.

The action implies EM which generalize those of Frenkel and Bargmann-Michel-Telegdi to the case of an arbitrary
electromagnetic field. They have been studied in \cite{DPM3}, \cite{deriglazov2014Monster}. For the present discussion
we need only symbolic form of the equation for position variable, this reads
\begin{eqnarray}\label{FF.3}
\frac{d}{d\tau}\left[\frac{\dot x^\mu}{\sqrt{-\dot xg\dot x}}\right]=f^\mu\,,
\end{eqnarray}
where $f^\mu$ is the polynomial $f^\mu=\sqrt{-\dot xg\dot x}a_1+a_2+\frac{a_3}{\sqrt{-\dot xg\dot x}}+\frac{a_4}{\dot
xg\dot x}$, with the coefficients ${a_i}$ that are finite as $\dot xg\dot x\rightarrow 0$. The effective metric arises
for the particle with anomalous magnetic moment $\mu\ne 1$
\begin{eqnarray}\label{FF.4}
g_{\mu\nu}=[\eta+b(\mu-1)(JF+FJ)+b^ 2(\mu-1)^2FJJF]_{\mu\nu},
\end{eqnarray}
where $b\equiv\frac{-2e}{4m^2c^3-3e\mu(JF)}$. The variational problem (\ref{FF.1}) yields also the value-of-spin and
Pirani \cite{pirani:1956, dixon:1964, tulczyjew:1959}
 conditions
\begin{eqnarray}\label{FF.5}
J^{\mu\nu}J_{\mu\nu}=6\hbar^2, \quad J^{\mu\nu}{\cal P}_\nu=0,
\end{eqnarray}
where the canonical momentum is ${\cal P}_\nu=\frac{\partial L}{\partial\dot x^\nu}-\frac{e}{c}A_\nu$. They provide
consistent quantization which yields one-particle (positive energy) sector of the Dirac equation, see \cite{DPM2}.

Inclusion of these constraints into a variational problem, as well as the search for an interaction consistent with
them turn out to be rather non trivial tasks \cite{hanson1974relativistic, mukunda1982}, and the expression
(\ref{FF.1}) is probably the only solution of the problem. So, the appearance  of effective metric (\ref{FF.4}) in Eq.
(\ref{FF.3}) seems to be unavoidable in a systematically constructed model of spinning particle.

The speed of light does not represent special point of the manifestly
relativistic equation (\ref{FF.3}). Singularity occurs at the critical velocity ${\bf v}_{cr}$ determined by the
equation $\dot xg\dot x=0$. We analyze this equation for the case of stationary homogeneous field, $\partial_\alpha
F_{\mu\nu}=0$. Using the consequence
%\begin{eqnarray}\label{L.14.1}
$(\dot xJF\dot x)=-b(\mu-1)(\dot xFJJF\dot x)$
%\end{eqnarray}
of the Pirani condition, and the expression $J^{\mu}{}_\alpha
J^{\alpha\nu}=-4\left[\pi^2\omega^\mu\omega^\nu+\omega^2\pi^\mu\pi^\nu\right]$, we write
\begin{eqnarray}\label{FF.6}
-(\dot xg\dot x)=-\dot x^2+4b^2(\mu-1)^2\left[\pi^2(\omega F\dot
x)^2+\omega^2(\pi F\dot x)^2\right].
\end{eqnarray}
As $\pi$ and $\omega$ are space-like vectors \cite{DPM3}, the last term is non-negative, so $|{\bf v}_{cr}|\ge c$. Let
us confirm that generally this term is nonvanishing function of velocity, then $|{\bf v}_{cr}|> c$. Assume the
contrary, that this term vanishes at some velocity, then
\begin{eqnarray}\label{FF.7}
\omega F\dot x=-\omega^0({\bf E}{\bf v})+({\boldsymbol{\omega}}, c{\bf E}+{\bf v}\times{\bf B})=0\,, \cr \pi F\dot
x=-\pi^0({\bf E}{\bf v})+({\boldsymbol{\pi}}, c{\bf E}+{\bf v}\times{\bf B})=0\,.
\end{eqnarray}
This implies $c({\bf D}{\bf E})+({\bf D}, {\bf v}\times{\bf B})=0$. Consider the case ${\bf B}=0$, then it should be
$({\bf D}{\bf E})=0$. On other hand, for the homogeneous field the quantity $J^{\mu\nu}F_{\mu\nu}=2\left[({\bf D}{\bf
E})+2({\bf S}{\bf B})\right]=2({\bf D}{\bf E})$ is a constant of motion \cite{DPM3}. Hence we can take the initial
conditions for spin such, that $({\bf D}{\bf E})\ne 0$ at any instant.

Thus, the critical speed does not always coincide with the speed of light and, in general case, we expect that ${\bf
v}_{cr}$ is both field and spin-dependent quantity. Using spacial part of Eq. (\ref{FF.3}), we can estimate
acceleration near $v_{cr}$. Explicit computation gives ${\bf v}{\bf a}\sim\sqrt{-\dot x g\dot x}$, that is the
acceleration along the direction of velocity vanishes as $|{\bf v}|\rightarrow v_{cr}>c$.

To see, whether we can keep the condition $v_{cr}=c$, we
use formal similarity of the matrix $g$ appeared in (\ref{FF.3}) with space-time metric. So, let us stop for a moment
to discuss the definitions of velocity and acceleration in general relativity. Consider pseudo Riemann space
\begin{eqnarray}\label{La.1}
{\bf M}^{(1,3)}=\{x^\mu,  ~  g_{\mu\nu}(x^\rho),  ~  g_{00}<0\}.
\end{eqnarray}
Physical velocity and acceleration can be defined in such a way that speed of light represents an observer independent
quantity and, more over, a particle during its evolution in curved background can not exceed the speed of light.

%We define three-dimensional velocity and acceleration which guarantee $c$ to be
%invariant scale of general relativity, as well as the equality $v_{cr}=c$.
To achieve this, we represent interval in $1+3$ block-diagonal form \cite{Landau:2}
\begin{eqnarray}\label{La.2}
-ds^2=g_{\mu\nu}dx^\mu dx^\nu= \qquad \qquad \qquad \cr
-c^2\left[\frac{\sqrt{-g_{00}}}{c}(dx^0+\frac{g_{0i}}{g_{00}}dx^i)\right]^2+\left(g_{ij}-\frac{g_{0i}g_{0j}}{g_{00}}\right)dx^idx^j.
\end{eqnarray}
This prompts to introduce infinitesimal time interval, distance and speed as follows:
\begin{eqnarray}\label{La.3}
dt=\frac{\sqrt{-g_{00}}}{c}(dx^0+\frac{g_{0i}}{g_{00}}dx^i), \qquad \quad \cr
dl^2=(g_{ij}-\frac{g_{0i}g_{0j}}{g_{00}})dx^idx^j\equiv\gamma_{ij}dx^idx^j, \quad
v=\frac{dl}{dt}.
\end{eqnarray}
The conversion factor between the world time $x^0$ and the physical time $t$ is
\begin{eqnarray}\label{La.3.1}
\frac{dt}{dx^0}=\frac{\sqrt{-g_{00}}}{c}(1+\frac{g_{0i}}{g_{00}}\frac{dx^i}{dx^0}).
\end{eqnarray}
Introduce also the three-velocity vector
\begin{eqnarray}\label{La.5}
v^i=\left(\frac{dt}{dx^0}\right)^{-1}\frac{dx^i}{dx^0},
\end{eqnarray}
or, symbolically $v^i=\frac{dx^i}{dt}$. This is consistent with the above definition of $v$:
$v^2=\left(\frac{dl}{dt}\right)^2={\bf v}^2=v^i\gamma_{ij}v^j$. In the result, the interval (\ref{La.2}) acquires the
form similar to special relativity (but now we have ${\bf v}^2={\bf v}\gamma{\bf v}$)
\begin{eqnarray}\label{La.6}
-ds^2=-c^2dt^2+dl^2=-c^2dt^2\left(1-\frac{{\bf v}^2}{c^2}\right).
\end{eqnarray}
This equality holds in any coordinate system $x^\mu$. Hence all observers conclude that a particle with $ds^2=0$ has
the speed ${\bf v}^2=c^2$.

To proceed further, we associate with ${\bf M}^{(1,3)}$ the one-parameter family of three-dimensional spaces ${\bf
M}^3_{x^0}=\{x^k, ~ \gamma_{ij}, ~  D_k\gamma_{ij}=0\}$. The covariant derivative $D_k$ is constructed with help of
three-dimensional metric $\gamma_{ij}(x^0, x^k)$ written in Eq. (\ref{La.3}), where $x^0$ is considered as a parameter
(below we also use $D_0=\frac{dx^ k}{dx^0}D_k$). Note that velocity has been defined above with help of the curve in
${\bf M}^3_{x^0}$ parameterized by this parameter, $x^i(x^0)$.

To define an acceleration, we need to adopt some notion of a constant vector field (parallel transport equation) in
this space. In Euclidean space the scalar product of two constant fields has the same value at any point. In
particular, taking the scalar product along a line $x^i(x^0)$, we have $\frac{d}{dx^0}(\xi, \eta)=0$. For the constant
fields in our case it is natural to demand the same (necessary) condition: $\frac{d}{dx^0}[\xi^i(x^0)\gamma_{ij}(x^0,
x^i(x^0))\eta^i(x^0)]=0$. Taking into account that $D_k\gamma_{ij}=0$, this condition can be written as follows
\begin{eqnarray}\label{La.8.3}
(D_0\xi+\frac12(\xi\partial_0\gamma\gamma^{-1}), \eta)+(\xi, D_0\eta+\frac12(\gamma^{-1}\partial_0\gamma\eta))=0.
\nonumber
\end{eqnarray}
So we take the parallel-transport equation to be
\begin{eqnarray}\label{La.8.4}
D_0\xi^i+\frac12(\xi\partial_0\gamma\gamma^{-1})^i=0.
\end{eqnarray}
Then we define the acceleration with respect to physical time as follows:
\begin{eqnarray}\label{La.8.5}
a^i=\left(\frac{dt}{dx^0}\right)^{-1}\left[D_0v^i+\frac12({\bf v}\partial_0\gamma\gamma^{-1})^i\right].
\end{eqnarray}
For the special case of constant gravitational field, $g_{\mu\nu}(x^i)$, the definition (\ref{La.8.5}) reduces to that
of Landau and Lifshitz, see page 251 in \cite{Landau:2}.

Let us estimate the acceleration as $v\rightarrow c$. Particle in general relativity propagates along geodesics of
${\bf M}^{(1,3)}$, with the reparametrization-invariant equation being
%$\lambda$ (we denote  $\dot x^\mu=\frac{dx^\mu}{d\lambda}$)
\begin{eqnarray}\label{La.10}
\frac{d}{d\tau}\left( \frac{\dot x^\mu}{\sqrt{-\dot xg\dot x} } \right)+
\Gamma^\mu{}_{\alpha\beta}(g)\dot x^\alpha\frac{\dot x^\beta}{\sqrt{-\dot xg\dot x}}=0.
\end{eqnarray}
To see, which equation for $a^i$ implied by (\ref{La.10}), we take $\tau=x^0$ in its spacial part, and write the
resulting expression through $v^i$ and then through $a^i$. Straightforward computation gives the expression
\begin{eqnarray}\label{La.11}
a^i=\frac12\left(\frac{d\tau}{dx^0}\right)^{-1}\left[({\bf v}\partial_0\gamma\gamma^{-1})^i-\frac{({\bf
v}\partial_0\gamma{\bf v})}{c^2}v^i\right]+ \cr \left(\delta^i{}_j-\frac{v^i({\bf
v}\gamma)_j}{c^2}\right)[\tilde\Gamma^j{}_{kl}(\gamma)v^kv^l-\left(\frac{d\tau}{dx^0}\right)^{-2}\Gamma^j{}_{00}- \cr
\Gamma^j{}_{kl}v^kv^l-2\left(\frac{d\tau}{dx^0}\right)^{-1}\Gamma^j{}_{0k}v^k].
\end{eqnarray}
Then the acceleration along the velocity is
\begin{eqnarray}\label{La.20}
{\bf v}\gamma{\bf a}=\frac12\left(\frac{d\tau}{dx^0}\right)^{-1}\left[({\bf v}\partial_0\gamma{\bf v})-({\bf
v}\partial_0\gamma{\bf v})\frac{({\bf v}\gamma{\bf v})}{c^2}\right]+ \cr \left(1-\frac{{\bf v}\gamma{\bf
v}}{c^2}\right)({\bf
v}\gamma)_i[\tilde\Gamma^i{}_{kl}(\gamma)v^kv^l-\left(\frac{d\tau}{dx^0}\right)^{-2}\Gamma^i{}_{00}- \cr
\Gamma^i{}_{jk}v^jv^k-2\left(\frac{d\tau}{dx^0}\right)^{-1}\Gamma^i{}_{0k}v^k]. \qquad \quad
\end{eqnarray}
This implies ${\bf v}\gamma{\bf a}\rightarrow 0$ as ${\bf v}\gamma{\bf v}\rightarrow c^2$. That is acceleration along
the direction of velocity vanishes as the speed approximates the speed of light. With these definitions the spinless
particle in an external gravitational field can not overcome the speed of light.

The last term in the definition (\ref{La.8.5}) yields the important factor $({\bf v}\partial_0\gamma\gamma^{-1})^i$ in
Eq. (\ref{La.11}). As EM (\ref{La.11}) do not contain $\sqrt{c^2-{\bf v}\gamma{\bf v}}$, they have sense when
$v>v_{cr}$. Without this factor, the particle could exceed $c$ and then continues to accelerate.

The construction can be applied, without modifications, to the effective metric (\ref{FF.4}) appeared in EM of spinning
particle (\ref{FF.3}).  Adopting that velocity and acceleration of spinning particle in electromagnetic field are given
by Eqs. (\ref{La.5}) and (\ref{La.8.5}), we obtain the theory with critical speed equal to the speed of light.

Due to $\mu-1$\,-factor in Eq. (\ref{FF.4}), the deformation of world-line geometry in
electromagnetic field will be seen only by a particle with anomalous magnetic moment. In a gravitational field the
deformed geometry could be probed by any spinning particle. To see this, let us consider the Frenkel electron in a
curved background with the metric $\tilde g_{\mu\nu}$. We use the model constructed in \cite{DPW2}. The Lagrangian can
be obtained from (\ref{FF.1}) (with $A^\mu=0$) replacing $\eta_{\mu\nu}$ by $\tilde g_{\mu\nu}$ and usual derivative of
$\omega^\mu$ by the covariant derivative, $\dot\omega^\mu ~ \rightarrow ~
D\omega^\mu=\dot\omega^\mu+\Gamma^\mu_{\alpha\beta}(\tilde g)\dot x^\alpha\omega^\beta$. This leads to EM consistent
with those of Papapetrou \cite{Papapetrou:1951pa}, the latter are widely assumed as the reasonable equations of
spinning particle in gravitational fields \cite{Hojman:2013, Poplawski:2010, Shiromizu:1999, Chicone}. In the EM (\ref{FF.3}) now appears
the effective metric \cite{DPW2}
\begin{eqnarray}\label{La.20.1}
g_{\mu\nu}=\left[\tilde g+\beta (J\theta+\theta J)+\beta^2 \theta J J \theta \right]_{\mu\nu},
\end{eqnarray}
where $\beta \equiv \frac{1}{8m^2c^2 - J\cdot \theta}$, $\theta_{\mu\nu} \equiv R_{\mu\nu\alpha\beta}J^{\alpha\beta}$
and $R_{\mu\nu\alpha\beta}(\tilde g)$ is the curvature tensor. Hence, to guarantee the observer-independence of $c$ and
the equality $v_{cr}=c$, we need to define velocity (\ref{La.5}) and acceleration (\ref{La.8.5}) using the deformed
metric $g$ instead of $\tilde g$.

%\section{Acknowledgments}

{\bf Acknowledgments.} The work of AAD has been supported by the Brazilian foundation CNPq (Conselho Nacional de
Desenvolvimento Científico e Tecnológico - Brasil). WGR thanks CAPES for the financial support (Programm PNPD/2011).

\end{document}